\documentclass[useAMS,usenatbib,referee]{mn2e}
\usepackage{bm}
\input epsf.tex
\usepackage{graphicx}
\def\b{\bm}
\title[Kolmogorov Dissipation scales in Weakly Ionized Plasmas]
       {Kolmogorov Dissipation scales in Weakly Ionized Plasmas}

\author[V. Krishan and Z. Yoshida]{V. Krishan$^{1,2,3}$\thanks{E-mail:
vinod@iiap.res.in} and Z. Yoshida$^4$ \\
$^{1}$Indian Institute of Astrophysics, Bangalore 560034, India\\ 
$^{2}$Raman Research Institute, Bangalore 560080, India\\
$^{3}$Solar-Terrestrial Environment Laboratory, Nagoya University,Nagoya, Aichi, Japan\\
$^{4}$Graduate School of Frontier Sciences,The University of Tokyo\\
5-1-5 Kashiwanoha, Kashiwa-shi,Chiba 277-8561, Japan \\ }

\begin{document}

\date{Accepted--\hskip 1 cm   Received in original form --}

\pagerange{\pageref{firstpage}--\pageref{lastpage}} \pubyear{2008}

\maketitle

\label{firstpage}
\begin{abstract}
In a weakly ionized plasma, the evolution of the magnetic field is
described by a ``generalized Ohm's law'' that includes the Hall effect
and the ambipolar diffusion terms.  These terms introduce additional spatial and time scales which play a decisive role in the cascading and the dissipation mechanisms in magnetohydrodynamic turbulence. We determine the Kolmogorov dissipation scales for the
viscous, the resistive and the ambipolar dissipation mechanisms. The
plasma, depending on its properties and the energy injection rate, may
preferentially select one of the these dissipation scales. thus determining the shortest spatial scale of the supposedly self-similar spectral distribution of the magnetic field. The results
are illustrated taking the partially ionized part of the solar
atmosphere as an example. Thus the shortest spatial scale of the supposedly self-similar spectral distribution of the solar magnetic field is determined by any of the four dissipation scales given by the viscosity, the Spizer resistivity ( electron-ion collisions), the resistivity due to electron-neutral collisions and the ambipolar diffusivity. It is found that the ambipolar diffusion dominates for resonably large energy injection rate. The robustness of the magnetic helicity in the partially ionized solar atmosphere would facilitate the formation of self-organized vortical structures.

\end{abstract}
\begin{keywords}
Partially ionized plasma, Kolmogorov dissipation, Hall effect, Ambipolar diffusion. 
\end{keywords}

\maketitle

\section{Introduction}
It is widely recognized that some ideal integrals of motion (i.e.,
constants of motion in the dissipation-less limit) play an essential
role in the complex nonlinear dynamics leading to remarkable
self-organizing structures.  The ``helicity'' is one such quantity
that imposes a ``topological'' constraint on a field Moffat (1978).
Constancy of the helicity is not attributed to any ``symmetry'' of the
system, but is due to a ``defect'' (or a singularity) of the
Poisson-bracket operator, and, thus, is a robust constraint throughout
the evolution.

A dissipative mechanism can change the helicity to remove the
topological constraint.  Self-organization may be understood as a
subtle balance between the conservation (restriction) and the
dissipation (relaxation).

One of the most successful models of self-organization is
due to Taylor (1974), who invoked the constancy of the magnetic helicity
$H=\int \bm{A}\cdot\bm{B}~d^3x/2$ (the integral is taken
over the total volume).  Minimizing the magnetic energy $E=\int
|\bm{B}|^2~d^3x/2$ (he omitted the kinetic and the thermal energies)
for a fixed $H$ yields an Euler-Lagrange equation
$\nabla\times\bm{B}=\lambda\bm{B}$ ($\lambda$ is the Lagrange
multiplier for restricting $H$), whose solution is the ``Beltrami
field'' representing a twisted force-free magnetic field. The cross helicity $H_c=\int \bm{V}\cdot\bm{B}~d^3x/2$ is also an invariant of the ideal magnetohydrodynamic system with flows where $\bm V$ is the fluid velocity.

This ``variational principle'' is based on the assumption that the
energy $E$ is preferentially (selectively) dissipated while the
helicity $H$ is approximately conserved (and that no other conserved
quantity puts an obstacle for the minimization of $E$).  A possible
justification for this assumption is given by the ``energy cascade''
ansatz ---the energy density of the fluctuating field cascades toward
small spatial scales, and the variations (spatial derivatives) of the
field are enhanced.  Then, the dissipation of the energy, which
includes higher-order spatial derivatives in comparison with the
helicity, proceeds much faster than that of the helicity.
This ``scale change'' of fluctuations may be represented by the
cascade of the energy density in the Fourier (wave number) space.

It has been established that $E$ and $H$ are the integrals of motion
of an ideal magnetohydrodynamic system, the energy cascade yields the
selective dissipation of the energy (with respect to the helicity)
leading to the Taylor relaxed state.  This paper extends the scope of
these considerations to ``weakly ionized plasmas''. There are many astrophysical systems with a rather low degree of
ionization dominated by the charged particle-neutral collisions and
the neutral particle dynamics. A major part of the solar photosphere
(Leake and Arber 2006; Krishan and Varghese
2008;), the protoplanetary disks (Krishan and Yoshida
2006) and the molecular clouds (Brandenburg and Zweibel 1994) are some
of the examples of weakly ionized astrophysical plasmas. 
 In addition these systems are believed to be turbulent. The evolution of the magnetic fields in such a plasma would
be affected by the multifluid interactions in general and the
ambipolar diffusion in particular (Zweibel 1988). It is important to know
 which dissipation scale would be the most effective in a given
 situation.
 
 Here, We (1) set up the energy and the helicity evolution equations in a weakly ionized plasma and
(2)determine the Kolmogorov dissipation scales for the different
energy dissipation mechanisms. 
The weakly ionized part of the solar atmosphere is presented as an example to illustrate the
predominance of one or the other dissipation scale.
 We show that the ambipolar effect, which is nonlinearly enhanced when 
the energy injection rate is large, may dominate the energy dissipation,
while it conserves the helicity.

\section{Energy and helicities in weakly ionized plasmas}
The dynamics of a weakly ionized plasma can be described with the
following equations ( Krishan and Yoshida 2006, Krishan and Varghese 2008, Krishan and Gangadhara, 2008 ):
the momentum balance of
the neutral fluid is given by:
\begin{equation}
 \frac{\partial \bm{V}}{\partial t}+
(\bm{V} \cdot\nabla)\bm{V} =-\nabla h
+\frac{\bm{J}\times\bm{B}}{c\rho_n} + \mu \nabla^2 \bm{V} ,
\label{neutral-momentum-evolution}
\end{equation}
 and the magnetic field $\bm{B}$ evolves as:
\begin{equation}
\frac{\partial \bm{B}}{\partial t}
=\nabla\times\left\{\left[ \bm{V} 
-\frac{\bm{J}}{en_e}+\frac{\bm{J}\times\bm{B}}{c\nu_{in}\rho_i}
\right] \times\bm{B} \right\}
+\eta{\nabla}^{2}\bm{B} .
\label{magnetic-field-evolution}
\end{equation}
and
\begin{equation}
\nabla\times\bm{B}=\frac{4\pi}{c}\bm{J}
\end{equation}
where $\bm{V}$ is the velocity of the neutral fluid and $h$ is the
total enthalpy. The Lorentz force in the neutral dynamics appears due
to the ion- neutral coupling in that the Lorentz force on the ions ($en_i(\b E + \frac{\b V_i\times\b B}{c}$)) is
balanced by the ion-neutral collisional force ($-\rho_i\nu_{in}(\b V_i-\b V_n)$) where $\nu_{in}$ is the ion- neutral collision frequency. The ion inertial force
has been neglected in comparison with the ion-neutral frictional
force. The viscosity $\mu$ of the neutral fluid arises due to neutral-neutral collisions with frequency $\nu_{nn}$. The electron inertial force is also neglected. We have further
assumed a constant density incompressible system.
We may write the electric diffusivity $\eta$ as:
\begin{equation}
\eta = \frac{c^2}{4\pi} \frac{m_e \nu_{en}}{n_e e^2} = \delta_e^2 \nu_{en} ,
\label{eta-0}
\end{equation}
and the kinematic viscosity $\mu$ as:
\begin{equation}
\mu = \frac{v_{n,th}^2}{\nu_{nn}} = \lambda_{n}^2 \nu_{nn} ,
\label{mu-0}
\end{equation}
where $\delta_e := c/\omega_{pe} = c/\sqrt{4\pi n_e e^2/m_e}$ is the
electron inertial length, and $\lambda_n := v_{n,th}/\nu_{nn}$ is the
neutral mean free path: $\nu_{en}$, $\nu_{nn}$ and $\nu_{ei}$ are the
electron-neutral, the neutral-neutral, and the electron-ion collision frequencies;
$\nu_{ei}<< \nu_{en}$ in the case of the weakly ionized plasma.

Assuming constant densities and incompressibility, we normalize the
variables in the ``Alfv\'enic units''.  We chose a representative
neutral flow velocity $V_0$ and a magnetic field $B_0$ related as:
\begin{equation}
\frac{\rho_n V_0^2}{2} = \frac{B_0^2}{8\pi}.
\label{Alfven_unit}
\end{equation}
Solving (\ref{Alfven_unit}) for $V_0$ gives a virtual Alfv\'enic
velocity $V_0=B_0/(4\pi \rho_n)^{1/2}$. If $\rho_n$ were $\rho_i$, this
$V_0$ is the well-known Alfv\'en velocity.  In a weakly ionized
plasma,
\begin{equation}
\alpha:= \frac{\rho_n}{\rho_i} \gg 1
\label{alpha}
\end{equation}
For example, in the solar atmosphere, $\alpha$ can be as large as
$10^{6}$, the protoplanetary disks have $\alpha$ of the order of
$10^8$.

Let $L_0$ be a characteristic length in the system.  We normalize 
$\bm{x}$ by $L_0$, $t$ by $t_0:=L_0/V_0$, $\bm{B}$ by $B_0$, $\bm{V}$
by $V_0$ and energy densities by $\rho_nV_0^2=B_0^2/(8\pi)$.  Using
these variables, Equations (1) and (2) can be written in the
(\ref{neutral-momentum-evolution}) and
(\ref{magnetic-field-evolution}) dimension-less form:
\begin{eqnarray}
& &
\frac{\partial \bm{V}}{\partial t}+
(\bm{V} \cdot\nabla)\bm{V} =-\nabla h
+\bm{J}\times\bm{B} + \epsilon_{\mu} \nabla^2 \bm{V} ,
\label{neutral-momentum-evolution-0}
\\
& & \frac{\partial \bm{B}}{\partial t}
=\nabla\times\left\{ \left[ \bm{V} 
-\epsilon_H \bm{J} + \epsilon_A \bm{J}\times\bm{B}\right]\times\bm{B} \right\}
 + \epsilon_{\eta}{\nabla}^{2}\bm{B} ,
\label{magnetic-field-evolution-0}
\end{eqnarray}
where the scaling parameters are defined as:
\begin{eqnarray}
\epsilon_H &:=& \alpha \frac{c/\omega_{pi}}{L_0}  = \alpha \frac{\delta_i}{L_0},
\label{Hall_term}
\\
\epsilon_A &:=& \epsilon_H \frac{\omega_{ci}}{\nu_{in}},
\label{ambipolar_term}
\\
\epsilon_{\eta} &:=&  \eta \frac{t_0}{L_0^2} = \eta \frac{1}{L_0 V_0},
\label{resisitvity-0}
\\
\epsilon_{\mu} &:=& \mu \frac{t_0}{L_0^2} = \mu \frac{1}{L_0 V_0} ,
\label{viscosity-0}
\end{eqnarray}
Here, $\delta_i$ is the ion inertial length.  The scaling parameter
$\epsilon_H$ multiplying the Hall term is enhanced by the factor
$\alpha = \rho_n/\rho_i$ in comparison with the standard (fully
ionized) Hall term.
 

With an appropriate homogeneous boundary conditions, we have the
energy equation ($E=\int (V^2 + B^2) dx/2$),
\begin{equation}
\frac{dE}{dt}= - \epsilon_A \int J_\perp^2 B^2~d^3x -\epsilon_{\eta}\int J^2~d^3x 
-\epsilon_{\mu}\int |\nabla\times\bm{V}|^2~d^3 x,
\label{Energy}
\end{equation}
and, for completeness, the helicity equation
($H=\int \bm{A}\cdot\bm{B} dx/2$)
\begin{equation}
\frac{dH}{dt} = -\epsilon_{\eta} \int \bm{J}\cdot\bm{B}~d^3x .
\label{Helicity}
\end{equation}


In the ideal limit ($\epsilon_A=0, \epsilon_{\eta} = 0$ and
$\epsilon_{\mu}=0$), $E$ and $H$ are conserved.  The constancy of $H$ is
destroyed only by a finite resistivity $\epsilon_{\eta}$. The cross helicity $H_c$ of the ideal MHD transforms to the  ion canonical helicity  $H_G=\int (\bm{A}+\epsilon_H\bm {V})\cdot\nabla\times(\bm{A}+\epsilon_H\bm {V}) dx/2$. The cross helicity $H_c$ is conserved when $\epsilon_H=0, \epsilon_A=0$ and  the ion helicity $H_G$ is conserved for $\epsilon_A=0$. For $\epsilon_A\neq 0$ both $H_c$ and $H_G$ are not conserved. From
(\ref{Energy}), we find that the energy dissipation is contributed by
(i) ambipolar diffusion (scaled by $\epsilon_A$), (ii) resistivity
(friction of electrons with neutrals; scaled by $\epsilon_{\eta}$), and
(iii) neutral viscosity (scaled by $\epsilon_{\mu}$).The advective terms $(\bm V\cdot\nabla)\bm V$,
 $\nabla\times (\bm V\times\bm B)$ and $\nabla\times (\bm J\times\bm B)$ ( the Hall term ) are responsible for the energy cascade mechanism. Since the velocity and the magnetic field are coupled through equations (8) and (9), all the advective and the dissipative processes operate on both the velocity and the magnetic fields.
We shall define the Reynolds number in a rather broad sense as the ratio of the advective term and the dissipation term. 
  We will see that the Reynolds numbers resulting from different combinations of the advective process and the dissipation process set up a rather complex scale
hierarchy of the Kolmogorov microscale, in the wave number ($k$) space, which changes depending on
the strength of the energy injection rate. In what follows we will select one dissipation mechanism and combine it with the three advective ( cascading) mechanisms to define the three Reynolds numbers and determine the corresponding Kolmogorov microscales.



\section{Scale Hierarchy}

The wave-number ($k$) space is divided into (i) the energy injection
(large scale) range, (ii) the ``inertial range'', and (iii) the
dissipation range.  The inertial range is dominated by the convective
[$(\bm{V}\cdot\nabla)\bm{V}$], inductive
[$\nabla\times(\bm{V}\times\bm{B})$] and the Hall
[$-\nabla\times(\epsilon_H\bm{J}\times\bm{B})$] effects, which create
a sub-hierarchy in the inertial range.  The higher-$k$-end of the
inertial range is the ``Kolmogorov scale'' that is determined by one
of the three (ambipolar, viscous and resistive) dissipation
mechanisms.  The aim of this section is to estimate the Kolmogorov
scale by identifying the responsible (i.e., the dominant) mechanism of
energy dissipation.

Here we invoke Kolmogorov's ansatz of ``local interactions'' in the
$k$-space, which may be formulated as follows.  Let $K$ symbolize a
``range'' of wave-number space, i.e., $K$ stands for $\{ \bm{k};~ K
\leq |\bm{k}| < K + \Delta \}$, where $\bm{k}$ is the wave vector and
$\Delta$ is a certain positive constant.  Denoting by
$\hat{u}(\bm{k})$ the Fourier transform of a field $u(\bm{x})$, we
define $u_K = \sum_{\bm{k}\in K}
\hat{u}(\bm{k})e^{i\bm{k}\cdot\bm{x}}$, which means the component of
$u(\bm{x})$ in the hierarchy of wave-numbers ranging in $K$.  Suppose
a term $X$ is included in an evolution equation of a physical quantity
$u$.  When we observe the hierarchy $K$, $X_K$ contributes the
temporal variation $(\partial u/\partial t)_K$.  If $X$ is a linear
term including a field $v$ (and, possibly, the differential operator
$\nabla$), then we may estimate $X_K$ only by $v_K$.  But, if $X$
includes, for example, $v\cdot w$, then all $\hat{v}(\bm{k}_1)$ and
$\hat{w}(\bm{k}_2)$ (and the corresponding $\nabla$ translating into
$i\bm{k}_1$ and $i\bm{k}_2$) may contribute to $X_K$, if $\bm{k}_1 \pm
\bm{k}_2 \in K$.  Now, the ``locality ansatz'' claims that only $v_K$
and $w_K$ (and, thus, $\nabla$ of order $K$) dominates $X_K$.  We
assume that this ansatz holds for all (even higher-order) nonlinear
terms in our system.  We also assume that no geometric anisotropy
diminishes the magnitudes of vector products such as
$\bm{J}\times\bm{B}$.  Hence, we estimate, for example,
$|\bm{J}\times\bm{B}|_K \approx |\bm{J}|_K |\bm{B}|_K \approx K
|\bm{B}|_K^2$ (using $\bm{J}=\nabla\times\bm{B}$ in the normalized
unit).  In what follows, we denote $|\bm{B}|_K$ by $B_K$.

\subsection{Kolmogorov scales defined by viscosity dissipation}

Using a rather general definition of the Reynolds number as the ratio of the term responsible for the cascade process to that responsible for the dissipation process, we consider, first, the advective  term $(\bm V\cdot\nabla)\bm V$ and the dissipation due to  the viscosity. This defines the very familiar, the (standard) Kolmogorov scale, when the other mechanisms (ambipolar
diffusion and resistivity) are still negligible.

The Reynolds number is defined by
\begin{equation}
R_\mu (K) := \frac{|(\bm{V}\cdot\nabla)\bm{V}|_K}{|\epsilon_{\mu}\nabla^2\bm{V}|_K} .
\label{Reynolds-0}
\end{equation}
Note that this Reynolds number is evaluated for each scale hierarchy
as a function of $K$.  Invoking the previous assumptions, we estimate
\begin{equation}
R_\mu (K) \approx \frac{V_K}{\epsilon_{\mu} K } .
\label{Reynolds-0-approx}
\end{equation}
The ``Kolmogorov microscale'' is usually defined as the spatial scale $l$ at which the Reynolds number becomes unity ( e.g. Holmes, Lumley and Berkoz 1996, Frisch 1995 ). We prefer to work in the wavevector space such that $K=l^{-1}$ and define the inverse Kolmogorov microscale  $K=K_\mu$ where $R_\mu(K)$ 
becomes of the order of unity ( e.g. Pope 2000). 
We obtain
\begin{equation}
K_\mu = \frac{V_{K_\mu}}{\epsilon_{\mu}} .
\label{Kolmogorov-0}
\end{equation}
To estimate (eliminate) $V_{K_\mu}$, we invoke the energy cascade rate ( total energy $E$ being an invariant of the system)
$\cal E$ ( normalized by ${\cal E}_0 = V^3_0/L_0$ ) that is assumed to be scale invariant and equal to the
energy dissipation rate as well as the energy injection rate. From Eq.(13)
\begin{equation}
{\cal E} = \epsilon_{\mu} K_\mu^2 V_{K_\mu}^2 
= \frac{V_{K_\mu}^4}{\epsilon_{\mu}}  .
\label{energy-dissipation-rate-1}
\end{equation}
Solving (\ref{energy-dissipation-rate-1}) for $V_{K_\mu}$, we obtain
\begin{equation}
V_{K_\mu} = {\cal E}^{1/4} {\epsilon_{\mu}}^{1/4} .
\label{Kolmogorov-scale-V}
\end{equation}
Substituting (\ref{Kolmogorov-scale-V}) into (\ref{Kolmogorov-0}) yields
the well-known relation
\begin{equation}
K_\mu = {\cal E}^{1/4}{\epsilon_{\mu}}^{-3/4}  .
\label{Kolmogorov-1}
\end{equation}
If the energy cascade is dominated by the advective term $\nabla\times (\bm V\times\bm B)$ and the dissipation by the viscosity; the
 Reynolds number $R_{\mu M}(K)$ should better represent the relation
between the energy cascade and the dissipation where

\begin{equation}
R_{\mu M}(K) :=\frac{|\nabla\times(\bm{V}\times\bm{B})|_K}{\epsilon_{\mu}|\nabla^2\bm{V}|_K}
\approx \frac{B_K}{\epsilon_{\mu} K }
= \frac{V_K}{\epsilon_{\mu} K C_{V/B}} ,
\label{Reynolds-m}
\end{equation}
where we have introduced a coefficient $C_{V/B} := V_K/B_K$ that is,
in general, a function of $K$.  The corresponding Kolmogorov scale
becomes
\begin{equation}
K_{\mu M} = \frac{B_{K_{\mu M}}}{\epsilon_{\mu}} = \frac{V_{K_{\mu M}}}{\epsilon_{\mu} C_{V/B}}
= {\cal E}^{1/4} {\epsilon_{\mu}}^{-3/4} C_{V/B}^{-1}.
\label{Kolmogorov-scale-B}
\end{equation}
In the ideal MHD regime, however, we may assume $C_{V/B} \approx 1$
throughout the MHD inertial range, because the energy transfer is an
equal collaboration of the induction and the convection terms.  Hence,
$K_{\mu M} \approx K_\mu$.

The situation changes, when the Hall term dominates the energy
cascade.  This is the case if the ion skin depth multiplied by the
density ratio $\alpha$, $\epsilon_H$ (normalized by the system size)
is larger than the Kolmogorov length scale, i.e., $\epsilon_H K_\mu >
1$.  One can define the Hall Reynolds number by taking the advective term  $\epsilon_H\bm J\times\bm B$ ( the Hall term) and the dissipative term to be due to the viscosity. The corresponding Reynolds number is defined as
\begin{equation}
R_{\mu H}(K) 
:=\frac{|\epsilon_H \nabla\times(\bm{J}\times\bm{B})|_K}{\epsilon_{\mu}|\nabla^2\bm{V}|_K}
\approx \frac{\epsilon_H B_K^2}{\epsilon_{\mu} V_K }
= \frac{\epsilon_HV_K}{\epsilon_{\mu}  C_{V/B}^2 } 
\label{Reynolds-h}
\end{equation}
scales the ratio of the energy cascade rate and the dissipation.
Assuming $C_{V/B} = (\epsilon_H K)^p$ (for $K>1/\epsilon_H$) 
with a certain exponent $p$ 
(the Hall-MHD turbulence theory, Krishan and Mahajan 2005, predicts,
for R-mode turbulence $p=1$, while for L-mode turbulence $p=-1$),
we obtain a ``Hall Kolmogorov scale'' (denoting $q:= 1/(2p)$)
\begin{equation}
K_{\mu H} = \epsilon_H^{q-1}\left( \frac{V_{K_{\mu H}}}{\epsilon_{\mu}}\right)^q
= \epsilon_H^{q-1} \left({\cal E}^{1/4}{\epsilon_{ \mu}}^{-3/4} \right)^q 
= \left(\epsilon_H^{-1}\right)^{1-q} \left(K_\mu \right)^q .
\label{Kolmogorov-scale-H}
\end{equation}
Since $K_{\mu H}/K_\mu = 1/(\epsilon_H K_\mu)^{1-q}$ 
(and we are assuming $\epsilon_H K_\mu >1$),
$K_{\mu H}$ interpolates $\epsilon_H^{-1}$ and $K_\mu$ (i.e.,
$\epsilon_H^{-1} < K_{\mu H} < K_\mu$) as long as $q<1$.  

As mentioned above, the condition for the appearance of the Hall MHD
regime is
\begin{equation}
\epsilon_H K_\mu = \epsilon_H {\epsilon_{\mu}}^{-3/4} {\cal E}^{1/4} > 1.
\label{Hall-MHD_condition}
\end{equation}

\subsection{Kolmogorov scales defined by resistivity dissipation}
Next, we examine the case where the resistive dissipation dominates over the viscous and the ambipolar diffusivities.
  Both
the viscous and the resistive dissipations have a common mathematical
structure viz. they are linear terms multiplied by $\nabla^2$.

The resistive and the viscous dissipation mechanisms may be compared as follows: from
(\ref{Energy}), the ratio of the resistive and viscous dissipation terms is
\begin{equation}
\frac{|\epsilon_{\eta} \nabla\times\bm{B}|^2_K}{|\epsilon_{\mu} \nabla\times\bm{V}|^2_K} =
\frac{\epsilon_{\eta} K^2 B_K^2}{\epsilon_{\mu} K^2 V_K^2} = \frac{\epsilon_{\eta}}{\epsilon_{\mu}} C_{V/B}^{-2}.
\label{ratio-mu-eta}
\end{equation}

Again taking $(\bm V\cdot\nabla)\bm V)$ as the advective term, in the MHD regime where we may assume $C_{V/B}\approx 1$, the
resistive Kolmogorov scale is given by just replacing
$\epsilon_{\mu}$ by $\epsilon_{\eta}$ in (\ref{Kolmogorov-1}), i.e.,
\begin{equation}
K_\eta = {\cal E}^{1/4} {\epsilon_{\eta}}^{-3/4} ,
\label{Kolmogorov-eta}
\end{equation}
where the energy dissipation rate is  ${\cal E} =\epsilon_{ \eta} K_\eta^2
B_{K_\eta}^2$.  Evaluating ${\cal E}$ for the same energy injection
rate, we may write $K_\eta = (\epsilon_{\eta} /\epsilon_{\mu})^{-3/4} K_\mu$.
By (\ref{eta-0}) and (\ref{mu-0}), we estimate
\begin{equation}
\frac{\epsilon_{\eta}}{\epsilon_{\mu}}
=  \frac{\delta_e^2 \nu_{en}}{\lambda_n^2\nu_{nn}} .
\label{eta_by_mu}
\end{equation}
For the MHD regime where we may assume $C_{V/B}\approx 1$ the advective term $\nabla\times (\bm V\times\bm B)$ along with the resistive dissipation furnishes the same scale $K_\eta$.

In the Hall MHD regime ($\epsilon_H K_\eta > 1$), we replace the
previous Hall Reynolds number (\ref{Reynolds-h}) by
\begin{equation}
R_{\eta H}(K) 
:=\frac{|\epsilon_H \nabla\times(\bm{J}\times\bm{B})|_K}{|\epsilon_{\eta}\nabla^2\bm{B}|_K}
\approx \frac{\epsilon_H B_K}{\epsilon_{\eta}}  .
\label{Reynolds-h-eta}
\end{equation}
Hence, at the corresponding Kolmogorov scale $K_{\eta H}$, we estimate
$B_{K_{\eta H}} = \epsilon_{\eta}/ \epsilon_H$.  From the energy dissipation rate
${\cal E} = \epsilon_{\eta} K_{\eta H}^2 B_{K_{\eta H}}^2$, we obtain
\begin{equation}
K_{\eta H} = {\cal E}^{1/2} \epsilon_{{\eta}}^{-3/2} \epsilon_H
= K_\eta^2 \epsilon_H .
\label{Kolmogorov-H-eta}
\end{equation}
Hence, unlike the relation of the viscosity Kolmogorov scales (\ref{Kolmogorov-scale-H}),
$K_{\eta H} > K_\eta$.

\subsection{Kolmogorov scales defined by ambipolar diffusion}
Finally, we estimate the Kolmogorov scales assuming that the dissipation is
dominated by the ambipolar term.

We begin with the MHD regime where we may assume $C_{V/B}\approx 1$. As shown earlier in this case the advective terms $(\bm V\cdot\nabla)\bm V)$ and $\nabla\times (\bm V\times\bm B)$ furnish equal Kolmogorov microscales.
We define an ``ambipolar Reynolds number'' by
\begin{equation}
R_A(K) := 
\frac{|\nabla\times(\bm{V}\times\bm{B})|_K}
{|\epsilon_A\nabla\times[ (\bm{J}\times\bm{B})\times\bm{B}]|_K}
\approx
\frac{V_K }{\epsilon_A K B_K^2}.
\label{Reynolds-A}
\end{equation}
The ambipolar Kolmogorov scale $K_A$ is characterized by $R_A(K_A) \approx 1$:
\begin{equation}
K_A = \frac{V_{K_A}}{\epsilon_A B_{K_A}^2}
= \frac{C_{V/B}}{\epsilon_A B_{K_A}} .
\label{Kolmogorov_scale-A}
\end{equation}
The ambipolar dissipation rate is now ( Eq. 13):
\begin{equation}
{\cal E} = \epsilon_A K_A^2 B_{K_A}^4 = \epsilon_A^{-1} C_{V/B}^2 B_{K_A}^2 .
\label{energy-dissipation-rate-A}
\end{equation}
Plugging $B_{K_A} = {\cal E}^{1/2} \epsilon_A^{1/2} C_{V/B}^{-1}$ into 
(\ref{Kolmogorov_scale-A}), and assuming $C_{V/B} \approx 1$, we obtain
\begin{equation}
K_A =  \epsilon_A^{-3/2} {\cal E}^{-1/2} .
\label{Kolmogorov_scale-A'}
\end{equation}

Let us compare (\ref{Kolmogorov_scale-A'}) with (\ref{Kolmogorov-1}):
\begin{equation}
\frac{K_A}{K_\mu} = 
\epsilon_{\mu}^{3/4} \epsilon_A^{-3/2} {\cal E}^{-3/4} .
\label{ratio-1}
\end{equation}
From this relation, we see that the ambipolar diffusion dominates over
the viscous dissipation (i.e., $K_A/K_\mu \ll 1$), when the energy
dissipation rate (=energy injection rate) ${\cal E}$ is sufficiently
large.  This is because the ambipolar dissipation is a nonlinear term
that is enhanced when the fluctuation level is high.

If the Hall term dominates the energy cascade i.e. taking $\epsilon_H \nabla\times(\bm{J}\times\bm{B}$ as the advective term
($ \epsilon_H K_A > 1$), we replace
(\ref{Reynolds-A}) by
\begin{equation}
R_{AH}(K) := 
\frac{|\epsilon_H \nabla\times(\bm{J}\times\bm{B})|_K}
{|\epsilon_A\nabla\times[ (\bm{J}\times\bm{B})\times\bm{B}]|_K}
\approx
\frac{\epsilon_H }{\epsilon_A  B_K}.
\label{Reynolds-AH}
\end{equation}
Using (\ref{energy-dissipation-rate-A}), we estimate
$B_{K_{AH}} = ({\cal E}/\epsilon_A)^{1/4} K_{AH}^{-1/2}$, which, together with
$R_{AH}(K_{AH}) =1$ yields
\begin{equation}
K_{AH} = \epsilon_A^{3/2} \epsilon_H^{-2} {\cal E}^{1/2}
= (\epsilon_H K_A)^{-2} K_A.
\label{Kolmogorov_scale-AH}
\end{equation}
We, thus, see $K_{AH} < K_A$.

 Comparing with (\ref{Kolmogorov-scale-H}), we get:
\begin{equation}
\frac{K_{AH}}{K_{\mu H}} = \epsilon_A^{3/2} 
\epsilon_H^{-(q+1)} \epsilon_{{\mu}}^{3q/4} {\cal E}^{(2-q)/4} 
= 
\left(\frac{\omega_{ci}}{\nu_{in}}\right)^{3/2}
\epsilon_H^{(1/2)-q} \epsilon_{{\mu}}^{3q/4} {\cal E}^{(2-q)/4}  .
\label{ratio-2}
\end{equation}

\subsection{Comparison of different Kolmogorov scales}

In Table~\ref{summary}, we summarize different estimates of Kolmogorov
scales.  When the plasma parameters and the energy injection rate
${\cal E}$ are specified, we can estimate the appropriate Kolmogorov
scale determined by the balance between the relevant energy cascade
and the dissipation mechanisms.

\begin{table}
\begin{center}
\begin{tabular}{c|l|l}
dissipation mechanism & MHD regime ($\epsilon_H K<1$)& H MHD regime ($\epsilon_H K>1$)\\
\hline
viscosity & $K_\mu ={\cal E}^{1/4} \epsilon_{{\mu}}^{-3/4}$ & $K_{\mu H} =(\epsilon_H^{-1})^{1-q}K_\mu^q ~~~(< K_\mu)$ \\
resistivity & $K_\eta ={\cal E}^{1/4} \epsilon_{{\eta}}^{-3/4}$ & $K_{\eta H} =\epsilon_HK_\eta^2 ~~~(> K_\eta)$ \\
ambipolar & $K_A ={\cal E}^{-1/2} \epsilon_A^{-3/2}$ & $K_{A H} =\epsilon_H^{-2}K_A^{-1} ~~~(< K_A)$  \\
\end{tabular}
\caption[Kolmogorov scales]{Summary of Kolmogorov scales}
\label{summary}
\end{center}
\end{table}

\section{Scale Hierarchy on the solar atmosphere}
The solar
magnetic flux, believed to be generated in the convection zone, has to pass through
the partially ionized solar photosphere before it can appear high up
in the solar corona. This realization is rather recent and is now
receiving a lot of attention. Arber, Haynes and Leake (2007) has
emphasized the profound effects on the temperature and the current
structure of the overlying chromosphere and the corona that the
inclusion of the neutral medium can produce. It is important to know which
dissipation scale would be the most effective for the conditions typical of the solar atmosphere. We
estimate the various Kolmogorov scales in the partially ionized part
of the solar atmosphere. 

The various collision frequencies are
determined from Khodochenko et al.(2004):
\begin{equation}
\nu_{ij}=\Sigma_{ij}n_n\left(\frac{8K_BT}{\pi m_{ij}}\right)^{1/2},
\label{}
\end{equation}
where (i,j) stands for the species of particles, the cross-section
$\Sigma_{en}\approx 10^{-15}\textrm{cm}^2 $, $\Sigma_{in}\approx \Sigma_{nn}\approx
5\times 10^{-15}\textrm{cm}^2 $, $m_{ij}=m_im_j\left(m_i+m_j\right)^{-1}$, $n_n$ is
the neutral particle density and $T$ is temperature in degree Kelvin. The typical values of the physical parameters on the solar atmosphere are given in Table II.

\begin{table}
\begin{center}
		\begin{tabular}{llllll}
\hline
$ h$ & $T$ & $\rho_i$ & $\rho_n$ & $B$ & $B_e$ \\
\hline 
0.   &   6520.& 1.00$\times10^{-10}$& 1.90$\times10^{-07}$& 1200.00& 9.4$\times10^{-04}$\\
50.  &   5790.& 1.20$\times10^{-11}$& 1.59$\times10^{-07}$& 1125.77& 1.2$\times10^{-04}$\\
125. &   5270.& 1.18$\times10^{-12}$& 1.00$\times10^{-07}$&  980.16& 1.3$\times10^{-05}$\\
175. &   5060.& 3.39$\times10^{-13}$& 7.04$\times10^{-08}$&  880.33& 4.6$\times10^{-06}$\\
250. &   4880.& 9.37$\times10^{-14}$& 3.89$\times10^{-08}$&  737.21& 1.7$\times10^{-06}$\\
400. &   4560.& 1.12$\times10^{-14}$& 1.09$\times10^{-08}$&  503.71& 4.2$\times10^{-07}$\\
490. &   4410.& 4.37$\times10^{-15}$& 4.84$\times10^{-09}$&  394.42& 2.6$\times10^{-07}$\\
560. &   4430.& 4.72$\times10^{-15}$& 2.47$\times10^{-09}$&  322.27& 4.2$\times10^{-07}$\\
650. &   4750.& 2.29$\times10^{-14}$& 1.00$\times10^{-09}$&  246.31& 3.7$\times10^{-06}$\\
775. &   5280.& 1.08$\times10^{-13}$& 3.79$\times10^{-10}$&  183.67& 3.5$\times10^{-05}$\\
855. &   5650.& 1.75$\times10^{-13}$& 1.66$\times10^{-10}$&  143.40& 1.0$\times10^{-04}$\\
980. &   5900.& 1.78$\times10^{-13}$& 6.57$\times10^{-11}$&  108.65& 1.8$\times10^{-04}$\\
1065.&   6040.& 1.67$\times10^{-13}$& 3.60$\times10^{-11}$&   90.88& 2.5$\times10^{-04}$\\
\hline
\end{tabular}
\end{center}
\end{table}

The variation of the ionization fraction $\frac{\rho_i}{\rho_n}$ with height on the solar atmosphere is shown in figure (1). 
\begin{figure*}
\includegraphics[width=10cm]{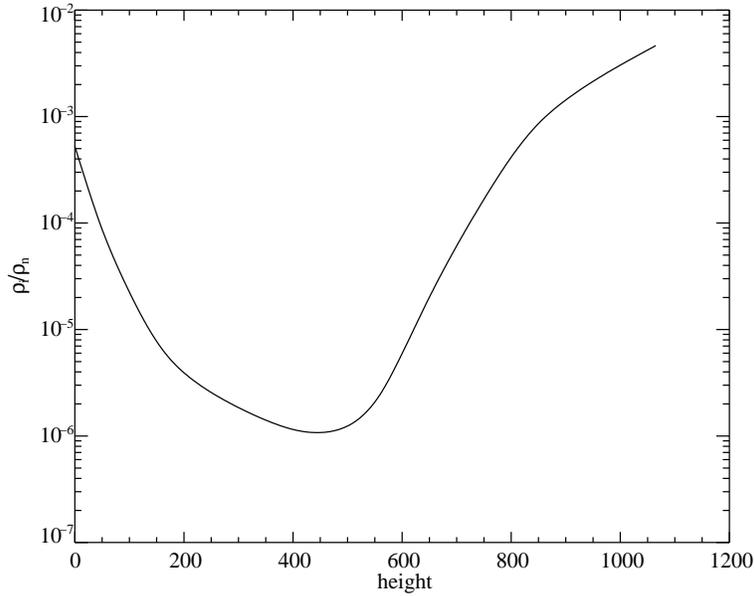}
\caption{ Variation of the Ionization fraction $\frac{\rho_i}{\rho_n}$ with height $h$ on the solar atmosphere. }
\end{figure*}

The ratios of the different Kolmogorov scales can be expressed as:
 \begin{equation}
\frac{K_A}{K_{\mu}}=3\times 10^{18}{\cal E}^{-3/4}\rho_i^{9/4}\rho_n^{-3/4}T^{9/8}B^{-3/2},\label{}
\end{equation}
\begin{equation}
\frac{K_{\eta}}{K_{\mu}}=10^{-7}\rho_i^{3/4}\rho_n^{-3/2}\label{}
\end{equation}

We present plots of the ratios
$\frac{K_A}{K_{\mu}}$, $\frac{K_{\eta}}{K_{\mu}}$ and $\frac{K_A}{K_{\eta}}$ vs height for different values of the energy injection rate ${\cal E}$ vs height in Figs. (2), (3) and (4) respectively.
\begin{figure*}
\includegraphics[width=10cm]{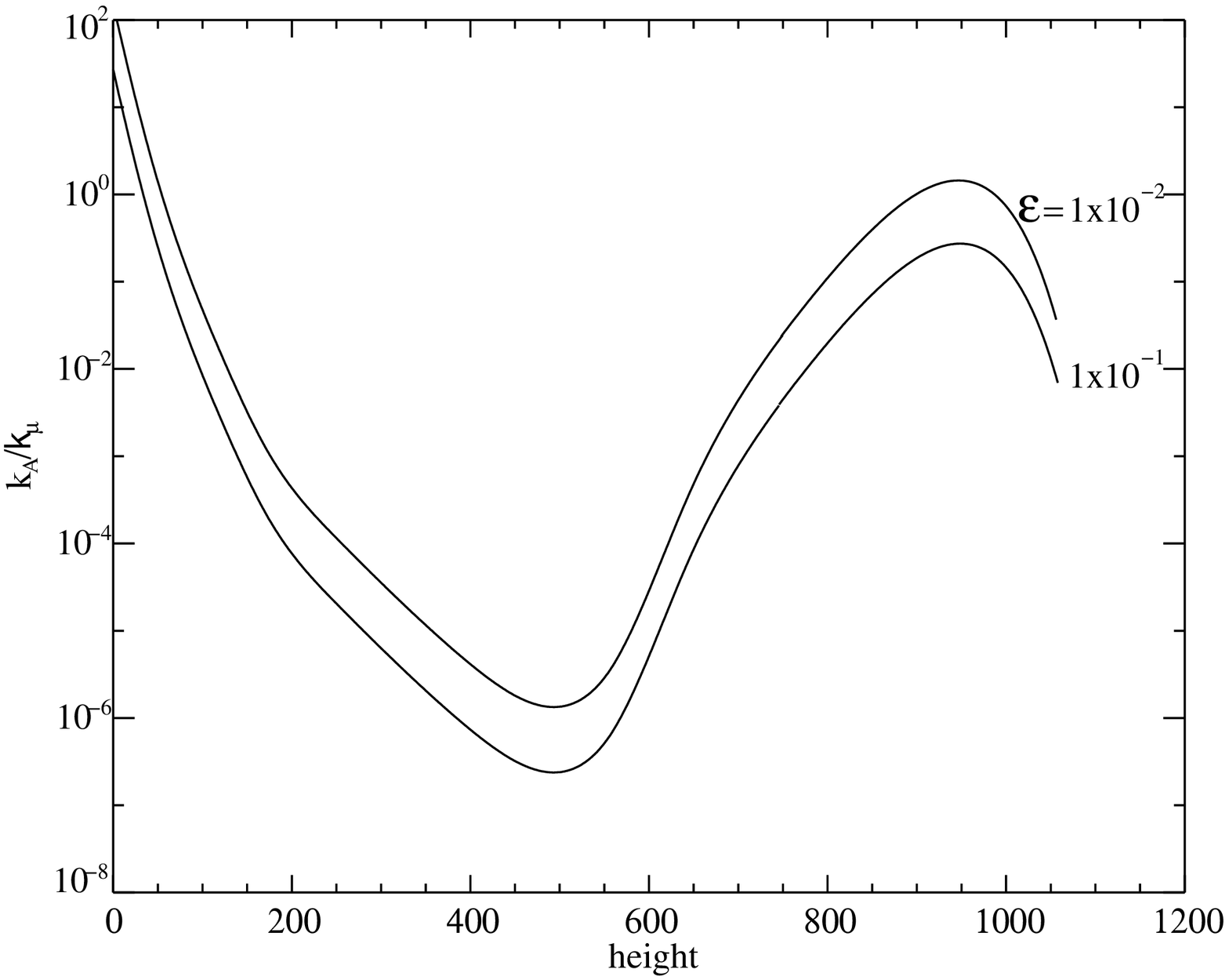}
\caption{ Variation of the ratio $\frac{K_A}{K_{\mu}}$ with height $h$ for different values of the energy injection rate ${\cal E}$  }
\end{figure*}
\begin{figure*}
\includegraphics[width=10cm]{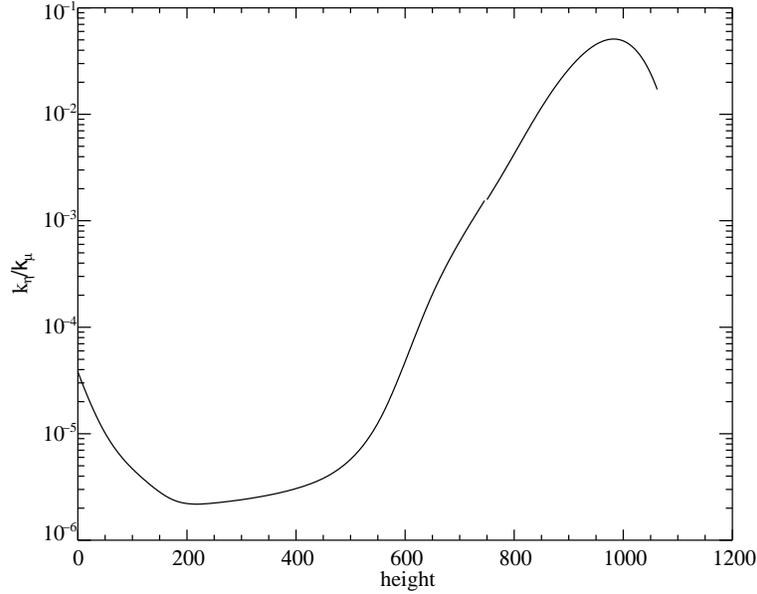}
\caption{ Variation of the ratio $\frac{K_{\eta}}{K_{\mu}}$ with height $h$ }
\end{figure*}
\begin{figure*}
\includegraphics[width=10cm]{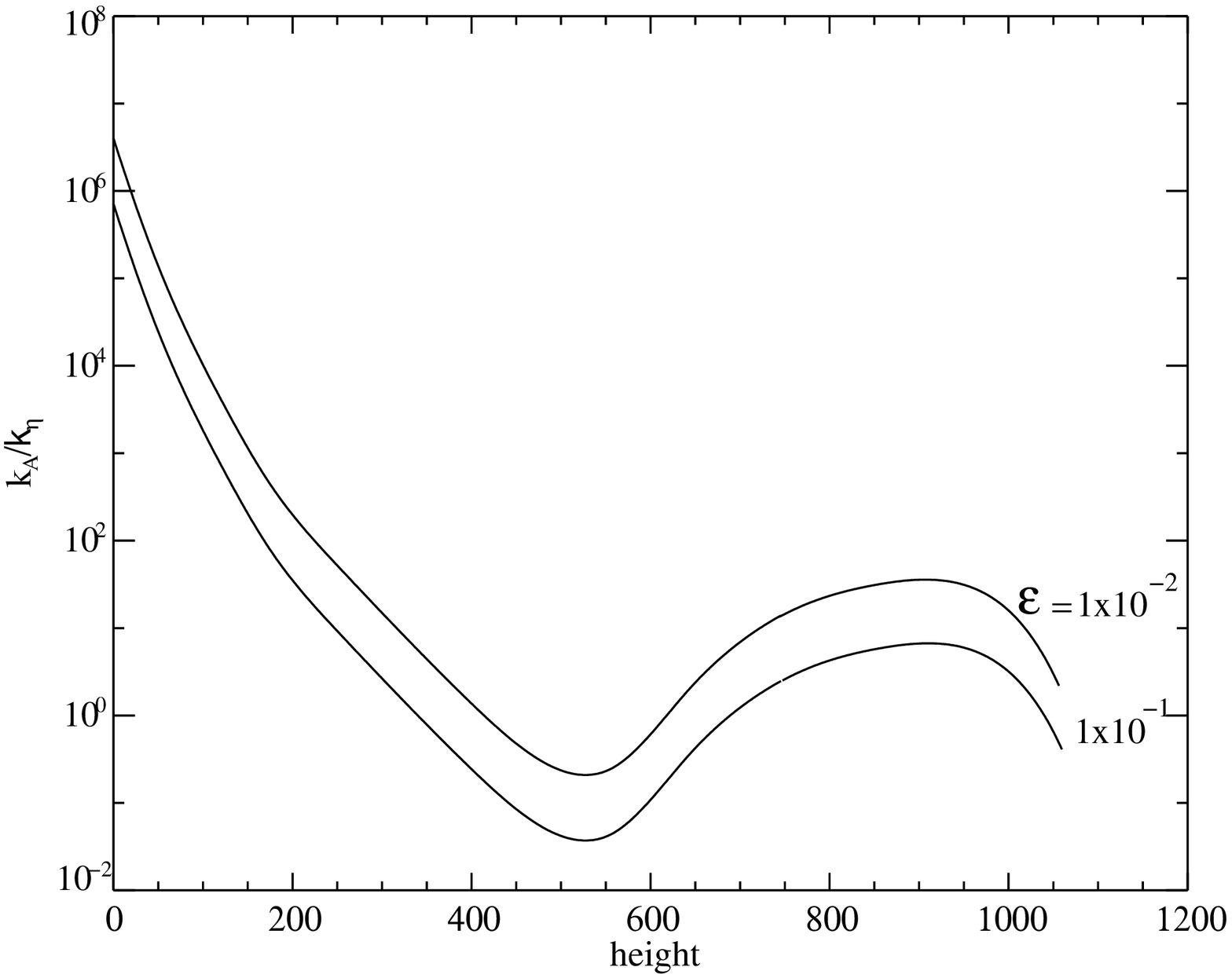}
\caption{ Variation of the ratio $\frac{K_A}{K_{\eta}}$ with height $h$ for different values of the energy injection rate ${\cal E}$ }
\end{figure*}

We
observe from Fig.(2) that the ambipolar dissipation predominates over the viscous
dissipation i.e. $K_A < K_{\mu}$ on major part of the solar photosphere and the chromosphere. Choosing the parameters at a height of $980$ Km above the
photosphere as the normalizing parameters i.e. $\rho_n = 6.5\times 10^{-11}$ g {cm}$^{-3}$, $B= 108$ G, $L_0\approx 1000$ Km, we find $V_0\approx 3.5$ Km s$^{-1}$ and ${\cal E_0}= 4\times 10^8$ cm$^2$ s$^{-3}$ . The condition for the predominance of the ambipolar diffusion over the viscous dissipation becomes
 ${\cal E} > 1.2\times 10^{-2}$ or, recovering dimensions, $5\times 10^6 cm^2 s^{-3}$. One can estimate the typical injection rate of the turbulent convective energy on the sun to be $V^3_c/L_0\approx 10^8$ $cm^2 s^{-3}$ by taking a typical convective velocity $V_c\approx 2 Km s^{-1}$. The condition for the predominance of the ambipolar diffusion is, thus, easily satisfied. The inclusion of the Hall effect modifies $K_A$ to $K_{AH}$ and as seen from table I $ K_{AH} < K_A$, reconfirming the predominance of the ambipolar effect as the dissipation mechanism. 
Figure (3) demonstrates the predominance of the resistive dissipation (due to electron-neutral collisions) over the viscous dissipation.   

That the ambipolar dissipation predominates over the resistive dissipation i.e. $K_A < K_{\eta}$ for the region extending from a height of 200 Km to 1000 Km  for reasonable values of the energy injection rate $\cal E$ can be seen in figure (4).
The Spitzer resistivity in a weakly ionized plasma, by definition, is much smaller than the resistivity due to the  electron-neutral collisions. 
We conclude that the ambipolar dissipation would dominate over other
dissipation mechanisms on most of the partially ionized region of the
solar atmosphere. 
\section{Conclusion}
The dynamics of weakly ionized plasmas is governed by the neutral fluid being subjected to the Lorentz force and the magnetic induction to the Hall and and the ambipolar effects. There are now three mechanisms by which the energy could dissipate, each mechanism having its own characteristic dissipation scale. We have determined these scales as presented in Table I. The example of the solar atmosphere shows that the ambipolar dissipation would determine the short scale end of the supposedly self-similar distribution of the photospheric magnetic field thereby identifying the region of dissipation leading to heating.
We observe that $H$ is a more robust quantity in comparison with $E$,
if either the ambipolar or the viscous dissipation is larger than the
resistive dissipation.  In the standard argument of selective
dissipation, moreover, even the resistivity dissipates $E$ faster than
it does $H$ because of the energy cascade toward small scales;
$dE/dt$ includes higher-order spatial derivatives in comparison
with $dH/dt$, so that $dE/dt$ assumes a larger value for small-scale
fields. This could lead to the formation of organized plasma structures, 
 vortical structures with twisted magnetic fields, 
so ubiquitous on the sun.
\section*{Acknowledgments}
The authors are grateful to Dr. A.B. Varghese for his help in the preparation of this manuscript.


\end{document}